\RequirePackage{fix-cm}
\documentclass[smallextended]{svjour3}       
\smartqed  
\usepackage{graphicx}
\newcommand{\be}{\begin{equation}}
\newcommand{\ee}{\end{equation}}
\newcommand{\bea}{\begin{eqnarray}}
\newcommand{\eea}{\end{eqnarray}}

\newcommand{\s}{\sigma}

\newcommand{\nn}{\nonumber\\}

\newcommand{\cL}{{\cal L}}
\newcommand{\br}{\bar R}
\newcommand{\bR}{\bar R}
\newcommand{\bg}{\bar g}

\newcommand{\bnabla}{\bar\nabla}

\newcommand{\tr}{{\rm tr}}

\newcommand{\Det}{{\rm Det}}

\newcommand{\diff}{\mathit{Diff}}
\newcommand{\sdiff}{\mathit{SDiff}}
\newcommand{\lich}{{\Delta_L}}

\begin{document}

\title{Unimodular quantum gravity\\ and the cosmological constant
}



\author{R. Percacci
}


\institute{R. Percacci \at
              SISSA, via Bonomea 256, 34134, Trieste, Italy \\
              and INFN, Sezione di Trieste, Trieste, Italy\\
              \email{percacci@sissa.it}           
}

\date{Received: date / Accepted: date}

\maketitle

\begin{abstract}
It is shown that the one-loop effective action of unimodular gravity 
is the same as that of ordinary gravity,
restricted to unimodular metrics.
The only difference is in the treatment of the global
scale degree of freedom and of the cosmological term.
A constant vacuum energy does not gravitate,
addressing one aspect of the cosmological constant problem.

\end{abstract}

\section{Introduction}
\label{intro}

The cosmological constant (CC) presents vexing issues to the
dominant worldview based on General Relativity (GR)
for the description of gravity
and Quantum Field Theory (QFT) for the description of matter fields
and their interactions.
The most striking way to present the problem is based on two assumptions:
first, that the observed acceleration of the universe is due
to a CC term in Einstein's equations
and second, that the energy density generated by vacuum fluctuations
depends quartically on a cutoff at the Planck scale. 
These assumptions lead to an 
apparent discrepancy of 123 orders of magnitude between 
the ``predicted'' and observed values of spacetime curvature.
Neither of these assumptions is based on very strong ground,
with the second being probably the weaker one.
For example, it has been argued that free massless fields
could have at most quadratic divergences
\cite{Akhmedov:2002ts,Ossola:2003ku}.
Furthermore, there exist ways of regulating the theory that 
completely avoid power divergences and will instead produce
an energy density proportional to the fourth power of
the masses of particles.
Taking typical electroweak masses reduces the discrepancy
to 55 orders of magnitude \cite{Shapiro:2000dz}, 
which is enormously smaller but still enormous.
We will refer to this as the first CC problem.

A second problem is that the CC receives contributions
from the vacuum energy of the Higgs field and from the
quark condensate of QCD.
The vacuum energy is expected to change during the phase
transitions in the early universe
by an amount that is very large compared to its present value.
In order to arrive at the present value
a conspiracy involving all the known interactions
seems to be required.

Finally, a third problem is to explain why the CC has a value
that is roughly comparable to the current 
averaged matter energy density.
It is hard to derive such a value from microphysics
\cite{Weinberg:1988cp,Burgess:2013ara}.

A very simple and compelling solution to the first problem 
is given by Unimodular Gravity (UG)
\cite{Anderson:1971pn,Ng:1999tx,Finkelstein:2000pg,Buchmuller:1988wx,Ellis:2010uc}.
The solution comes from the fact that in UG a constant vacuum energy does not gravitate, and therefore does not contribute to spacetime curvature.
\footnote{Changes in the vacuum energy do affect the equations,
as we shall discuss below.}
We will review this in section 2.

One may then wonder what becomes of this argument
when one goes from the classical to the quantum theory.
This is relevant irrespective of the UV completion and even 
if gravity is an ``emergent'' phenomenon \cite{Percacci:2010af}.
Insofar as the CC problem is an infrared one,
one would expect that it can be solved within the
effective field theory description.

The differences between GR and UG have been discussed several times
in the literature.
At the classical level it is known that the two theories
are equivalent except for the status of the global scale of the metric,
and of the CC \cite{Henneaux:1989zc}.
Besides the equations of motion, the equivalence has been seen also
in the tree-level amplitudes
\cite{Burger:2015kie,Alvarez:2016uog}.
The possibility of discriminating between the two theories
has been discussed in \cite{Alvarez:2009ga}.
At the quantum level the situation is much less clear.
The issue has been addressed at the level of 
the semiclassical theory \cite{Fiol:2008vk},
nonlinear canonical quantization \cite{Smolin,Bufalo:2015wda},
linearized covariant theory \cite{Alvarez:2008zw,Alvarez:2015pla},
BRST quantization \cite{Alvarez,Upadhyay:2015fna}
and functional renormalization group 
\cite{Eichhorn,Benedetti:2015zsw,Saltas:2014cta}.
In particular it has been argued in \cite{Padilla:2014yea}
that the two theories must be equivalent also at the level of
the path integral, and that any lack of equivalence can be
attributed to different quantization procedures.
The calculation of ultraviolet divergences in UG,
akin to the classic calculations of \cite{thv,Christensen:1979iy},
had been done before in \cite{Alvarez:2008zw,Alvarez:2015pla,Alvarez},
with results that differed from GR.
On the other hand a more recent calculation 
of the gravitational contribution to the 
scalar scattering amplitude yielded
the same result for both theories \cite{Gonzalez-Martin:2017fwz}.

In section 3 we will repeat this calculation in a simpler setup
and find that, in a well-defined sense, the one-loop effective actions
are the same in the two theories,
supporting the claim of \cite{Padilla:2014yea}.
This part is based on reference \cite{dlop}, but uses a different
approach to the construction of the path integral.

\section{Classical UG}

In the following, by GR we mean any metric theory of gravity,
without restricting the form of the action.
By UG we mean a reformulation of GR
where the determinant of the metric is fixed a priori and
is not subjected to variation.
Let
$\omega\,\epsilon_{\mu_1\ldots\mu_d}\,dx^{\mu_1}\wedge\ldots\wedge dx^{\mu_d}$
be a fixed, non-dynamical volume form.
We impose that the metric satisfies 
\be
\sqrt{|g|}=\omega\ .
\label{constraint}
\ee
We do not make any assumption about the dimension of the
coordinates, so $\omega$ may carry dimensions.
If the coordinates have dimension of length,
then $\omega$ is dimensionless and without loss
of generality we may assume that it is equal to one.
This is the origin of the name ``unimodular'',
but we shall use the same term also when $\omega$
is a general scalar density of weight one.

The condition (\ref{constraint}) should be seen as a choice of gauge,
removing one local degree of freedom from the metric.
The total volume (when defined)
is diffeomorphism-invariant and therefore
the condition $\int d^dx\sqrt{|g|}=\int d^dx\,\omega$,
which follows from (\ref{constraint}),
is a genuine physical constraint.
Therefore UG can also be formulated
as GR with the constraint that the total volume is fixed.
Physically, the two theories thus differ by a single
quantum mechanical degree of freedom (not one per spacetime point)
\cite{Henneaux:1989zc}.

The condition (\ref{constraint}) breaks diffeomorphism invariance.
The residual gauge group is
the group $\sdiff$ of volume-preserving or ``special'' diffeomorphisms.
The Lie algebra of infinitesimal gauge transformations
are the transverse vectorfields, satisfying 
\be
\nabla_\mu v^\mu=0\ .
\label{transv}
\ee

Let $S_{GR}=S_{GR}^g+S_{GR}^m$
be an arbitrary $\diff$-invariant action for gravity and matter,
where
$$
S_{GR}^g(g)=\int d^dx\,\sqrt{|g|}\, \cL_g(g)
\qquad\mathrm{and}\qquad 
S_{GR}^m(\psi,g)=\int d^dx\,\sqrt{|g|}\, \cL_m(\psi,g)
$$
are the gravitational and matter actions respectively.
As specific examples one can think of the Hilbert action
\be
\cL_g=Z_NR\ ,\qquad 
Z_N=\frac{1}{16\pi G}
\label{ehaction}
\ee
and minimally coupled scalar
\be
\cL_m=-\frac{1}{2}g^{\mu\nu}\partial_\mu\phi\partial_\nu\phi
-V(\phi)
\ee
and Maxwell fields
\be
\cL_m=-\frac{1}{4}g^{\mu\nu}g^{\rho\sigma}F_{\mu\rho}F_{\nu\sigma}\ ,
\ee
but the following discussion is general.
We assume at first that $\cL_g$ does not contain a cosmological
term. The case when it does will be discussed later.

We define an action for unimodular gravity
$S_{UG}=S_{UG}^g+S_{UG}^m$
by simply replacing $\sqrt{|g|}$ by $\omega$ in $S_{GR}$:
$$
S_{UG}^g(g)=\int d^dx\,\omega\, \cL_g(g)
\ ;\qquad
S_{UG}^m(\psi,g)=\int d^dx\,\omega\, \cL_m(\psi,g)\ .
$$

We now discuss the mechanism by which the equations of motion of UG 
turn out to be the same as those of GR,
except for the CC.
In varying $S_{UG}$ one must impose 
$$
\delta \sqrt{|g|}
\equiv\frac{1}{2}g^{\mu\nu}\delta g_{\mu\nu}=0\ .
$$
This leads to the tracefree equations
\be
-\tilde E^{\mu\nu}+\frac{1}{d}g^{\mu\nu}\tilde E
=\frac{1}{2}\left(\tilde T^{\mu\nu}-\frac{1}{d}g^{\mu\nu}\tilde T\right)\ ,
\label{teom}
\ee
where
$$
\tilde E^{\alpha\beta}=\frac{1}{\omega}\frac{\delta S_{UG}^g}{\delta g_{\alpha\beta}}
\qquad
\mathrm{and}
\qquad
\tilde T^{\alpha\beta}=\frac{2}{\omega}\frac{\delta S_{UG}^m}{\delta g_{\alpha\beta}}
$$
is a kind of symmetric but not conserved energy-momentum tensor.

Normally the covariant conservation of the energy-momentum tensor
follows from the diffeomorphism invariance of $S_{GR}^m$.
From the $\sdiff$-invariance of $S_{UG}^m$,
due to the constraint (\ref{transv}),
there follows the weaker condition
$\nabla_\mu \tilde T^{\mu\nu}=\nabla^\nu \Sigma$,
where $\Sigma$ is some scalar field.
This suggests defining an ``improved''
energy-momentum tensor $\tilde T^{\mu\nu}-g^{\mu\nu}\Sigma$,
that will be conserved.
One can easily guess what $\Sigma$ is
by considering the energy-momentum tensors 
for the scalar and Maxwell fields:
\be
\tilde T^{\mu\nu}=\nabla^\mu\phi\nabla^\nu\phi
\qquad\mathrm{and}\qquad
\tilde T^{\mu\nu}=F^{\mu\rho}F^\nu{}_\rho\ .
\ee
Using the equations of motion (and the Bianchi identity)
one finds
\be
\nabla_\mu \tilde T^{\mu\nu}=-\nabla^\nu \cL_m\ .
\ee
It is clear that in both cases the term coming from the
variation of $\sqrt{|g|}$ is missing,
and this term is proportional to the Lagrangian density.
We will therefore define the symmetric, conserved energy-momentum tensor
\be
T_{\mu\nu}=\tilde T_{\mu\nu}+g_{\mu\nu}\cL_m\ .
\label{improve}
\ee
This just reconstructs the usual energy-momentum tensor used in GR:
\be
T_{\mu\nu}=\frac{2}{\sqrt{|g|}}\frac{\delta S_{GR}^m}{\delta g_{\mu\nu}}\ .
\ee
Since $\tilde T_{\mu\nu}$ and $T_{\mu\nu}$ have the same tracefree part,
we can replace $\tilde T_{\mu\nu}$ by $T_{\mu\nu}$
in the r.h.s. of (\ref{teom}):
\be
-\tilde E^{\mu\nu}+\frac{1}{d}g^{\mu\nu}\tilde E
=\frac{1}{2}\left(T^{\mu\nu}-\frac{1}{d}g^{\mu\nu}T\right)\ .
\label{teom2}
\ee
Next, consider the action $S_{GR}^g$. Its $\diff$-invariance
implies the identity
\be
\nabla_\mu\tilde E^{\mu\nu}+\frac{1}{2}\nabla^\nu\cL_g=0\ .
\label{gbi}
\ee
In the case of the Hilbert action, this is the familiar
contracted Bianchi identity, so we shall refer to this
as a ``generalized Bianchi identity''.
Acting with $\nabla_\mu$ on both sides of (\ref{teom2})
and using (\ref{gbi}) we obtain
\be
\nabla^\nu\left(\frac{1}{2}\cL_g+\frac{1}{d}\tilde E
+\frac{1}{2d}T\right)=0\ 
\ee
and therefore
\be
\frac{1}{2}\cL_g+\frac{1}{d}\tilde E
+\frac{1}{2d}T=Z_N\Lambda_1
\ee
where $\Lambda_1$ is an arbitrary integration constant 
and the prefactor $Z_N$ is conventional.
Using this in (\ref{teom2}) one obtains
\be
-\tilde E^{\mu\nu}-\frac{1}{2}g^{\mu\nu}\cL_g
+Z_N\Lambda_1 g^{\mu\nu}
=\frac{1}{2}T^{\mu\nu}\ .
\label{teom3}
\ee
This is exactly the equation that one would obtain by varying
the action $S_{GR}$, except for the term involving $\Lambda_1$.
For example, if $\cL_g$ has the Hilbert form (\ref{ehaction}),
$\tilde E^{\mu\nu}=-Z_N R^{\mu\nu}$ (up to a total derivative)
and we obtain Einstein's equations with an arbitrary
CC:
\be
R^{\mu\nu}-\frac{1}{2}g^{\mu\nu}R
+\Lambda_1 g^{\mu\nu}
=\frac{1}{2Z_N}T^{\mu\nu}\ .
\label{teom4}
\ee

So far we assumed that $\cL_g$ does not contain a cosmological term.
Now assume that it contains $-2Z_N\Lambda_0$.
The same CC $\Lambda_0$ then appears in the equations of motion of GR.
In UG this term merely contributes a field-independent
constant to the action
and therefore does not contribute to the equations of motion,
which contain again an arbitrary CC $\Lambda_1$,
unrelated to $\Lambda_0$.

In UG the value of the ``physical'' CC
(the one that appears in the equation of motion (\ref{teom3}))
is fixed in
the process of solving the equations of motion
and is independent of any cosmological term that may be present
in the action.
In this way the first of the CC problems is eliminated.

If the matter action contains a potential 
for a scalar field $-V(\phi)$,
the CC can be identified 
with the value of the potential at the minimum.
The potential appears in the equations of motion of UG 
through the combination $V(\phi)+2Z_N\Lambda_1$.
Since $\Lambda_1$ is arbitrary, 
the value of the potential at its minimum 
can be shifted arbitrarily
and does not have a physical meaning.
However, changes in the value of the potential 
due to a rolling scalar field, such
as occur during inflation, do affect the equations
\cite{Ellis:2013eqs}.

\section{Quantum UG}
\label{sec:1}

In this section we will define the one-loop path integral for UG
with the standard Hilbert Lagrangian (\ref{ehaction}) 
and contrast it with the path integral for GR
with the same Lagrangian.
The procedure followed here is slightly different
from that used in \cite{dlop}, but it leads to the same results.

There are several ways of defining classical UG,
and the definition of the quantum theory may depend
upon the formulation one starts with.
One common way of defining UG is to start from
the Hilbert action with the standard measure factor $\sqrt{|g|}$, 
and add to the action a term
$$
\int dx \lambda(\sqrt{|g|}-\omega)\ ,
$$
where $\lambda$ is a Lagrange multiplier.
This complicates the calculation of radiative corrections.
We shall seek a minimal formulation avoiding the
introduction of auxiliary fields.

In the calculation of quantum effects it is inevitable to use
the background field method, splitting the
metric $g_{\mu\nu}$
into a classical (background) part $\bg_{\mu\nu}$
and a quantum fluctuation $h_{\mu\nu}$.
The usual additive parametrization 
$g_{\mu\nu}=\bg_{\mu\nu}+h_{\mu\nu}$ is ill-suited to discuss UG
(and more generally the action of Weyl transformations,
and the dynamics of the conformal factor
\cite{Machado:2009ph,Percacci:2011uf,Codello:2012sn,Nieto:2017ddk}
).
We will instead follow \cite{Eichhorn}
and use the exponential parametrization
\footnote{General arguments in favor of the use of the exponential
parametrization in quantum gravity can be found in
\cite{pv1,nink}.}
\be
g_{\mu\nu}=\bg_{\mu\nu}(e^X)^\rho{}_\nu
\label{exppar}
\ee
demanding that $X^\rho{}_\nu=\bg^{\rho\sigma}h_{\sigma\nu}$ 
be traceless.
Although it may not be immediately obvious, 
by Taylor-expanding the exponential one sees that if $h_{\mu\nu}$
is symmetric, then also $g_{\mu\nu}$ is.
Imposing that the background is unimodular
\be
\sqrt{|\bg|}=\omega\ ,
\label{bgconstraint}
\ee
then automatically enforces (\ref{constraint}),
and tracelessness is a linear condition that one can
confidently impose on the quantum field,
without use of Lagrange multipliers.
With this parametrization, the path integral of UG will be an
integral over symmetric traceless tensors $h_{\mu\nu}$.

\subsection{The path integral.}

Before discussing UG we shall first present the path integral
of GR in a way that makes it easy to draw a comparison between the
two theories.
We start from a formal Euclidean path integral over metrics
\be
Z=\int d\mu\,e^{-S_{GR}(g)}\ ,
\ee
where $d\mu$ is a suitable measure and
$S$ is the Hilbert action (\ref{ehaction}).
Using the background field method, the integration measure
can be chosen so that $d\mu=(dh)$, where $h_{\mu\nu}$ 
is the fluctuation field in the exponential splitting (\ref{exppar}).
In contrast to UG, in GR it is not traceless.

We would like to split the path integral into an integral
over the gauge group and an integral over physical degrees of freedom.
Instead of using the standard Faddeev-Popov procedure,
we will use here the geometrical approach to functional
integration over geometries, that has been discussed
in \cite{Mazur:1989by,Bern:1990bh,Mottola:1995sj}.
We begin by describing the method in the context of GR,
then the modifications due to the unimodularity condition
will be easy to track.

The symmetric tensor $h_{\mu\nu}$ can be decomposed
\begin{equation}
h_{\mu\nu}=h^{\mathrm{TT}}_{\mu\nu}
+\bar{\nabla}_{\mu}\xi_{\nu}+\bar{\nabla}_{\nu}\xi_{\mu}
+\left(\bar{\nabla}_{\mu}\bar{\nabla}_{\nu}
-\frac{1}{d}\bar{g}_{\mu\nu}\bar{\nabla}^{2}\right)\sigma
+\frac{1}{d}\bar{g}_{\mu\nu}h\,,
\label{ap1}
\end{equation}
where
\begin{equation}
\bar{\nabla}^{\mu}h^{\mathrm{TT}}_{\mu\nu}=0\,,\,\,\,\, \bar{g}^{\mu\nu}h^{\mathrm{TT}}_{\mu\nu}=0
\,,\,\,\,\, \bar{\nabla}^{\mu}\xi_{\mu}=0\,,\,\,\,\,
h=\bar{g}^{\mu\nu}h_{\mu\nu}\,.
\label{ap2}
\end{equation}
The Jacobian of the change of variables 
$(dh)=(dh^{TT}d\xi d\sigma dh)J_1$
is
\be
\label{j1}
J_1=\det\left(\lich_1-\frac{2\bR}{d}\right)^{1/2}
\det\left(\lich_0\right)^{1/2}
\det\left(\lich_0-\frac{\bR}{d-1}\right)^{1/2}
\ee
where $\Delta_L$ are the Lichnerowicz Laplacians:
\bea
\lich_0 \phi &=& -\bnabla^2 \phi, \nn
\lich_1 A_\mu &=& -\bnabla^2 A_\mu + \bR_\mu{}^\rho A_\rho, \nn
\lich_2 h_{\mu\nu} &=& -\bnabla^2 h_{\mu\nu} 
+\bR_\mu{}^\rho h_{\rho\nu}
+ \bR_\nu{}^\rho h_{\mu\rho} 
-\bR_{\mu\rho\nu\s} h^{\rho\s} 
-\bR_{\mu\rho\nu\s} h^{\s\rho}\ .
\eea

Now consider an infinitesimal diffeomorphism $\epsilon^\mu$.
We can decompose the transformation
parameter $\epsilon^\mu$ in its longitudinal and transverse parts
(relative to the background metric):
\be
\label{qpar}
\epsilon^\mu=\eta^\mu+\bnabla_\mu\lich_0^{-\alpha}\phi\ ;
\qquad
\bar\nabla_\mu\eta^\mu=0\ .
\ee
We have inserted here an inverse power of the Laplacian
in the definition of the transformation parameter.
We will see that the following constructions
do not depend on this choice of parameterization of $\phi$.

We can then calculate the separate transformation properties
of the York variables under longitudinal and transverse
infinitesimal diffeomorphisms.
We have
\be
\label{deltaxi}
\delta_{\eta}\xi^\mu=\eta^\mu\ ;\qquad
\delta_\phi h=-2\lich_0^{1-\alpha}\phi\ ;\qquad
\delta_\phi\sigma=2\lich_0^{-\alpha}\phi\ ,
\ee
all other transformations being zero.
Note that $\sigma$ and $h$ are gauge-variant but the scalar combination
\be
\label{ess}
s=h+\lich_0\sigma
\ee
is invariant.

We expand the action around an Einstein background
\be
\bR_{\mu\nu}=\frac{\bR}{d}\bg_{\mu\nu}\ .
\ee
An Einstein metric automatically satisfies the tracefree part
of the Einstein equations with CC $\Lambda_0$.
The remaining trace equation is
\be
E\equiv\bR-\frac{2d}{d-2}\Lambda_0=0\ .
\ee

The Hessian of the Hilbert action, in exponential parametrization
and after York decomposition, is 
(see \cite{pv1,pereiraI,pereiraII} or section 5.4.6 in \cite{perbook2}):
\bea
S^{(2)}_{GR}&=&\frac{Z_N}{2}\!\int d^dx\,\sqrt{\bg}
\Bigg[
\frac{1}{2}h^{\mathrm{TT}}_{\mu\nu}
\left(\Delta_{L2}-\frac{2\bR}{d}
\right)h^{\mathrm{TT}\mu\nu}
\nn
&&
\hskip1cm
-\frac{(d-1)(d-2)}{2d^2}s\left(\Delta_{L0}
-\frac{\bR}{d-1}\right)s
-\frac{d-2}{4d}E h^2\Bigg]
\,.
\label{ehgf1}
\eea
Note that on-shell ($E=0$), it depends only on the
$\diff$-invariant variables $h^{TT}$ and $s$.
The last term is entirely absent in UG.

\subsection{One loop effective action of GR}

The one-loop path integral is
\be
Z_{GR}=e^{-S_{GR}(\bg)}\int(dh^{TT}d\xi d\sigma dh)
J_1e^{-S^{(2)}_{GR}(h^{TT},s;\bg)}\ .
\label{pigr}
\ee
The form (\ref{ehgf1}) suggest changing the variables in the path integral
to include $s$ instead of one of the gauge-variant variables
$\sigma$ or $h$.
We can choose to work either with $(h^{TT}_{\mu\nu},\xi_\mu,\sigma,s)$
or $(h^{TT}_{\mu\nu},\xi_\mu,s,h)$.

In the first case, we easily see from (\ref{ess}) 
that the Jacobian of this further transformation is one,
so in the measure
\be
\label{tr2}
(d\sigma dh)=
(d\sigma ds)
\det\left(\frac{\partial(\sigma,h)}{\partial(\sigma,s)}\right)
=(d\sigma ds)\ .
\ee

Now from (\ref{deltaxi}) we see that $\xi_\mu$ and $\sigma$
transform by a shift under infinitesimal diffeomorphisms.
Therefore we can identify these variables as coordinates
in the gauge orbit through the background metric,
while $h^{TT}$ and $s$ are gauge-invariant
coordinates in the quotient space.

More precisely, let us first take $\epsilon^\mu$ as coordinates
in (an infinitesimal neighborhood of the identity in) the
group of diffeomorphisms.
The integration measure on diffeomorphisms is $d\mu_\diff=(d\epsilon)$.
We want to pass to the coordinates $\eta$ and $\phi$
defined in (\ref{qpar}).
Applying to (\ref{qpar}) the same reasoning that led to (\ref{j1}) we have
\be
d\mu_\diff=(d\epsilon)=(d\eta d\phi)\det\lich_0^{\frac{1}{2}-\alpha}\ .
\label{mudiff}
\ee

We can now identify a neighborhood of the gauge orbit through
the background metric with a neighborhood of the identity in $\diff$.
Using (\ref{deltaxi}) we can replace $\xi_\mu$ and $\sigma$ 
with $\eta$ and $\phi$ respectively.
Then, up to irrelevant numerical constants,
\bea
(d\xi d\sigma)J_1&=&(d\eta d\phi)\det\lich_0^{-\alpha}J_1
\nn
&=&d\mu_\diff
\det\left(\lich_1-\frac{2R}{d}\right)^{1/2}
\det\left(\lich_0-\frac{\bR}{d-1}\right)^{1/2}\ ,
\label{aurora}
\eea
where one of the determinants in $J_1$ has been used to reconstruct
the measure (\ref{mudiff}).
Note that the result is independent of $\alpha$.
Using this in the path integral:
\bea
Z&=&V_\diff \times e^{-S_{GR}(\bg)}\det\left(\lich_1-\frac{2R}{d}\right)^{1/2}
\det\left(\lich_0-\frac{\bR}{d-1}\right)^{1/2}
\nn
&&
\hskip3cm
\times
\int(dh^{TT}ds)e^{-S^{(2)}_{GR}(h^{TT},s;\bg)}\ ,
\label{z2}
\eea
where $V_\diff=\int d\mu_\diff$ .
We have thus succeeded in factoring the volume of
the gauge group, and we can drop this numerical
factor from the path integral.

Evaluating the Gaussian integrals, we find that the
integral over $s$ cancels the scalar determinant,
so that \cite{Christensen:1979iy}
\be
\Gamma_{GR}\equiv-\log Z_{GR}=S_{GR}
+\frac{1}{2}\log\Det\left(\Delta_{L2}-\frac{2\br}{d}\right)
-\frac{1}{2}\log\Det\left(\Delta_{L1}-\frac{2\br}{d}\right)\ .
\label{z3}
\ee

If instead of $h$ we decided to eliminate $\sigma$,
in place of (\ref{tr2}) we would find
\be
\label{tr3}
(d\sigma dh)=
(ds dh)
\det\left(\frac{\partial(\sigma,h)}{\partial(s,h)}\right)
=(ds dh)\det\lich_0^{-1}\ .
\ee
Due to (\ref{deltaxi}), $h$ can be identified with
$-2\lich_0^{1-\alpha}\phi$, so, using (\ref{mudiff}),
\bea
(d\xi d\sigma dh)J_1&=&
(d\xi ds dh)\det\lich_0^{-1}J_1
=(ds d\eta d\phi)\det\lich_0^{-\alpha}J_1
\nn
&=&(ds)d\mu_\diff
\det\left(\lich_1-\frac{2R}{d}\right)^{1/2}
\det\left(\lich_0-\frac{4\Lambda}{d-1}\right)^{1/2}\!.
\eea
We are thus led again to (\ref{z2}) and (\ref{z3}).

To summarize, we see that
the fields $\xi_\mu$ and either $\sigma$ or $h$
can be used as coordinates in the gauge orbits,
leading to the factorization of the volume of the gauge group.
The Faddeev-Popov determinant is a piece of
the Jacobian of the coordinate transformation.

It is useful to note that in the parametrization
$(\eta_\mu,\phi)$, the volume $V_\diff$
contains a power of the determinant of the scalar Laplacian.
When presented this way, it is not clear that it can be treated as a
field-independent constant and dropped.
This follows instead from seeing it as an integral over the
differentially unconstrained field $\epsilon_\mu$.
Such (divergent) integrals can always be 
formally treated as being metric-independent.

\subsection{One-loop effective action of UG}

In UG, instead of (\ref{pigr}) we have a path integral
\be
Z_{UG}=e^{-S_{UG}(\bg)}\int(dh^{TT}d\xi d\sigma)J_1
e^{-S^{(2)}_{UG}(h^{TT},\sigma;\bg)}\ .
\label{ugpi}
\ee
The only difference with the path integral of GR is the absence 
of the integral over the trace $h$.
The Jacobian $J_1$ is the same as before.

Again we want to factor the volume of the gauge group,
which is now $\sdiff$.
There are two observations to be made at this point.
First, since $h$ is not present, 
only the first of the two procedures discussed in the preceding sections
is available.
The variable $\sigma$ can be identified with $s$ (up to a Laplacian)
and is $\sdiff$-invariant.
The quadratic action reads
\bea
S^{(2)}_{UG}&=&\frac{Z_N}{2}\!\int d^dx\,\omega
\Bigg[
\frac{1}{2}h^{\mathrm{TT}}_{\mu\nu}
\left(\Delta_{L2}-\frac{2\bR}{d}
\right)h^{\mathrm{TT}\mu\nu}
\nn
&&\hskip2cm
-\frac{(d-1)(d-2)}{2d^2}
\sigma\lich_0^2\left(\lich_0
-\frac{\bR}{d-1}\right)\sigma
\Bigg]\,.
\label{ehgf2}
\eea
Second, the difference between the volume of $\sdiff$ and
the volume of $\diff$ lies in the absence of the integral over $\phi$
in (\ref{mudiff}).

Since the field $\eta_\mu$ is subject to the differential constraint
$\bnabla_\mu\eta^\mu=0$, which depends on the metric,
we cannot treat the formal measure $(d\eta)$ as field-independent,
as we do for $(d\epsilon)$.
The question then arises of how to split the determinant of $\lich_0$
in (\ref{mudiff}) between the integral over $\eta$ and the integral over $\phi$, in such a way that the two integrals can be treated
as metric-independent.

We determine the measure on the group $\sdiff$ by
matching the missing integral over $h$ in (\ref{ugpi})
to the missing integral over $\phi$ in the gauge group.
More precisely, we can identify the quotient space $\diff/\sdiff$,
infinitesimally parametrized by $\phi$, with the space of volume forms,
infinitesimally parametrized by $h$.
From (\ref{deltaxi}) we see that $(dh)=\det\lich_0^{1-\alpha}(d\phi)$,
so we split
$$
V_\diff=\int d\mu_\diff=
\int(d\eta)\det\lich_0^{-1/2}
\int(d\phi)\det\lich_0^{1-\alpha}
=V_\sdiff\int(dh) 
$$
where $V_\sdiff=\int d\mu_\sdiff$ and the measure 
on $\sdiff$ has been defined as
\be
d\mu_\sdiff=(d\eta)(\det\lich_0)^{-1/2}\ .
\label{elena}
\ee
According to the standard rules mentioned above,
both $V_\diff$ and $\int (dh)$ can be treated as metric-independent
and therefore so can $V_\sdiff$.
It is a factor that can be dropped from the functional integral.
\footnote{One can also get to the same result by the following argument
that was pointed out by D. Benedetti.
We define $V_\sdiff$ by inserting a delta function $\delta(\nabla_\mu\epsilon^\mu)$ in the integral over $\epsilon$.
Using (\ref{mudiff}) and $\nabla_\mu\epsilon^\mu=\nabla^2\phi$, we find
$V_\sdiff=\int(d\eta)(d\phi)\det\lich_0^{1/2}\delta(\lich_0\phi)
=\int(d\eta)(d\phi)\det\lich_0^{-1/2}\delta(\phi)=\int(d\eta)\det\lich_0^{1/2}$.
}

With this measure, instead of (\ref{aurora}) we have now
\bea
(d\xi)J_1&=&(d\eta)J_1
=d\mu_\sdiff\det\lich_0^{1/2}J_1
\nn
&=&d\mu_\sdiff\det\lich_0
\det\left(\lich_1-\frac{2R}{d}\right)^{1/2}
\det\left(\lich_0-\frac{\bR}{d-1}\right)^{1/2}.
\label{aurora2}
\eea
The integral over $\sigma$ gives
$$
\det\lich_0^{-1}\det\left(\lich_0-\frac{\bR}{d-1}\right)^{-1/2}
$$
and cancels the scalar determinants in (\ref{aurora2}).

The remaining Gaussian integrals work as before, 
so that $\Gamma_{UG}$ contains the same determinants as $\Gamma_{GR}$.
In particular, they will exhibit the same divergences.
We can think of using the same renormalization prescriptions for the
quadratic and logarithmic divergences.
However, in GR the volume (quartic) 
divergence has to be renormalized so as to
match the observed CC,
whereas in UG the cosmological term is field-independent and
can be dropped altogether, or renormalized to any other value.
We can therefore write for the renormalized effective actions
\be
\Gamma_{UG}=\Gamma_{GR}\Big|_{\sqrt{|g|}=\omega}+C\ ,
\label{result}
\ee
by which we mean that for any metric satisfying the
constraint (\ref{constraint}), the two functionals
are the same, up to an arbitrary constant.

By the general arguments given in section 2,
the quantum equations of motion derived from 
traceless variations of $\Gamma_{UG}$
will be the same as those obtained from unconstrained
variations of $\Gamma_{GR}$,
except that the value of the CC in the equations
is unrelated to the one appearing in the action.

\section{Discussion}

It has been shown in section 2 that, apart from the CC,
a general classical unimodular 
theory of gravity defined by a Lagrangian density $\cL$
is locally indistinguishable from classical GR 
with the same Lagrangian density.
This is a consequence of the fact that 
the scalar density $\sqrt{|g|}$ changes under $\diff$ and
can always be transformed locally into any predetermined function,
in particular a constant.
This means that given an arbitrary variation of the metric,
there exist infinitesimal diffeomorphisms
that can be used to cancel the effect 
of the variation on $\sqrt{|g|}$.
The discussion in section 3 makes this more explicit.
In the decomposition (\ref{ap1}) four variables are
pure gauge.
By identifying $\xi_\mu$ and $h$ with the coordinates of $\diff$,
we can eliminate $h$ from the physical degrees of freedom
and remain with a $\sdiff$-invariant unimodular theory.
The equations one gets from the traceless variations
of the metric yield the same equations of motion.
As we have seen, the solution of the traceless equations of motion
involves an arbitrary CC.
By decoupling the physical CC, which appears in the equations
of motion, from the CC, or vacuum energy, that may be present in
the Lagrangian, UG effectively solves the first of the
CC problems mentioned in the introduction.

This is the main reason for much of the interest in UG,
but it is worth recalling that there are at least two other
theoretical arguments in its favor.
The first, which was the starting point for 
the discussion in \cite{Ng:1999tx},
is of group-theoretic nature:
the spin-two representation of the Poincar\'e group
is contained in the traceless symmetric tensor.
Insofar as gravity can be seen as a force mediated by a
spin-two field, keeping also the trace part 
is completely unnecessary.

To state the second reason, let us recall that the physical
degrees of freedom of gauge theories are generally non-local
and unwieldy to work with.
It is mainly in order to work with local fields
that we accept the complications deriving from
the presence of an infinite-dimensional invariance group.
From this point of view it seems desirable to have as little 
gauge invariance as is needed to work with local variables.
\footnote{From a theoretical (as opposed to practical)
point of view, there are situations when the opposite is useful:
by increasing the number of gauge degrees of freedom
one can sometimes show that seemingly unrelated theories
are just different gauge-fixed versions of an overarching theory.
One example is the $GL(4)$-invariant reformulation
of gravity, that reduces to metric and tetrad formulations
in specific gauges
\cite{Percacci:2009ij,Floreanini:1989hq,Dabrowski:1986cf}.
Other examples where the gauge can be fixed without
generating non-localities are theories obtained
by the St\"uckelberg trick.}
In the case of GR (in four dimensions) one usually
works with ten local fields and a gauge group depending
on four functions, but this is not optimal.
Clearly UG is better on this count, 
because it can be formulated
in terms of nine local fields and a gauge group
depending only on three functions.
The reason why this works is that the unimodularity constraint
is algebraic rather than differential:
solving it leaves us with one less local degree of freedom,
and does not introduce any non-locality.

It is also interesting to note that the
well-known analogy between 
the classical theory of gravity
and the chiral models of particle physics
\cite{Percacci:2009ij}
also works better in UG.
In both cases the field is matrix-valued, subject to nonlinear constraints,
and the action can be expanded in powers of derivatives.
For the chiral models, the first terms in such an expansion are 
\bea
\label{chiral}
S&=&\int dx\left[L_2+L_4+O(\partial^6)\right]
\\
L_2&=&-\frac{F_\pi^2}{4}\tr(U^{-1}\partial U)^2
\nonumber\\
L_4&=&\ell_1 \tr(((U^{-1}\partial U)^2)^2)
+\ell_2(\tr(U^{-1}\partial U)^2)^2\ .
\nonumber
\eea
where $U$ has values in $SU(N)$.
In the case of gravity we have
\bea
S&=&\int dx\sqrt g\left[L_0+L_2+L_4+O(\partial^6)\right]
\\
L_0&=&m_P^2\Lambda
\nonumber\\
L_2&=&-\frac{1}{2}m_P^2R
\nonumber\\
L_4&=&\alpha R^2
+\beta R_{\mu\nu}R^{\mu\nu}
+\gamma R_{\mu\nu\rho\sigma}R^{\mu\nu\rho\sigma}
\nonumber
\eea
Recalling that the Christoffel symbols have the structure
${\mit\Gamma}\sim g^{-1}\partial g$ 
and that the curvature tensors contains terms
of the form ${\mit\Gamma}^2\sim (g^{-1}\partial g)^2$,
the analogy between the terms $L_2$ (and $L_4$) is striking.
In both cases, expanding the matrices around classical backgrounds leads
to very similar non-renormalizable perturbative expansions.

There appear to be two main differences between these actions:
in the gravitational case there is the nontrivial volume element
$\sqrt{|g|}$,
and there is a term $L_0$ containing no derivatives.
\footnote{Actually, in realistic applications also the chiral model
has a term without derivatives, but it contains different
powers of the field.}
We observe that in UG (especially if we choose $\omega=1$)
both of these differences are removed.
UG seems therefore to be an even better analog of the chiral models.

To summarize the main point so far,
we have seen that, aside from the total volume, 
which is fixed in UG and is a physical observable, 
the unimodularity constraint is merely a gauge condition.
Therefore, at the classical level, GR and UG are 
locally equivalent.

Passing now to the quantum theory, one may ask two questions.
The first is whether ``quantization preserves unimodularity''.
The second is whether the quantum theory constructed starting from 
the classical actions $S_{GR}$ and $S_{UG}$ ``are the same''.
One can make these questions precise in the quantum field theoretic
approach to quantum gravity, where one uses the background field method
to construct a covariant effective action that contains all the
effects due to the quantum fluctuations on the dynamics of the metric.
We have addressed these questions in a ``minimal'' formulation
where the metric of GR is seen as the exponential of a symmetric tensor
while the metric of UG is the exponential of a traceless symmetric tensor.
In this context, the first question is answered in the affirmative, 
because the exponential parametrization automatically enforces the
unimodularity constraint on the quantum field and its expectation value.
The second question also has a positive answer,
which is contained in equation (\ref{result}).

There have been conflicting claims in the literature on this last point.
In some cases the disagreements may simpy reflect different quantization procedures
or different classical starting points for the quantization procedure.
We believe that a quantization procedure that respects the classical
equivalence of UG and GR is to be preferred.
This is the case of the ``minimal'' formulation of UG
presented here.

The results reported here also hold for the formulation of UG
in which the metric is unconstrained, but there is
a Weyl symmetry in the action 
(sometimes referred to as ``$WT\diff$'' theory 
\cite{Alvarez:2006uu,Alvarez:2010cg,Bonifacio:2015rea,Oda:2016pok}).
The issue of the quantum equivalence of this formulation to GR
has been raised in \cite{pereiraI,pereiraII} 
and answered positively in \cite{dlop}.

These results also answer in the affirmative 
the question raised in \cite{Weinberg:1988cp}, 
whether UG can be obtained as the classical
limit of a satifactory quantum theory of gravity.
It is by now well-known that the effective field theory approach
to quantum gravity, pioneered in \cite{donoghue1},
provides a satisfactory (meaning: consistent and predictive)
description of gravity at sub-Planckian energies.
The same formalism may even have an UV completion based on the
same variables, if the asymptotic safety approach was successful
\cite{perbook2}.
The formalism developed in section 3 provides a unimodular
quantum theory of gravity having classical UG as a classical limit.
This certainly works at one loop level in the effective field theory.
It seems likely that the results can be generalized to higher loops
and there seems to be no obstacle to applying functional renormalization group methods to unimodular metrics,
possibly leading to an asymptotically safe
unimodular theory of gravity \cite{Eichhorn,Benedetti:2015zsw}.
We note that several results on GR that use the unimodular gauge
can be viewed as results for quantum UG
\cite{pv1,Labus:2015ska,opv,Dona:2015tnf}.

Whether it comes from an UV-complete theory, or just an effective 
field theory, the unimodular effective action $\Gamma_{UG}$
is just a classical action for an unimodular metric
and therefore we can apply to it the results of section 2.
In this way, the conclusion that
vacuum energy does not appear in the equations of motion 
is extended to the quantum domain.
This eliminates the ``prediction'' that spacetime should
have Planckian curvature, thus effectively defanging 
the first CC problem.
Attempts to use UG to explain away also the other
CC problems have been made in \cite{Ng:1999tx,Smolin}.

\begin{acknowledgements}
This paper is based in part on joint work with R. de Le\'on Ard\'on
and N. Ohta.
I also thank M. Henneaux and S. Gielen for discussions
and D. Benedetti and A. Eichhorn for reading parts of the manuscript
and making useful suggestions.

\end{acknowledgements}

\end{document}